\documentclass[aps,prl,twocolumn,groupedaddress,amsfonts,showpacs]{revtex4}
\usepackage{amsmath,amssymb,bm,graphicx,epsfig,psfrag,ulem}

\newcommand{\Rey}{\mbox{\rm Re}}            %Reynolds
\newcommand{\Rm}{R_\mathrm{m}}              %magnetic Reynolds
\newcommand{\pderiv}[2]{\frac{\partial #1}{\partial #2}} %Partial deriv

\begin{document}

\title{A Reconnecting Flux Rope Dynamo}

\author{Andrew~W.~Baggaley}
%\email{a.w.baggaley@ncl.ac.uk}
\author{Carlo~F.~Barenghi}
\author{Anvar~Shukurov}
\affiliation{School of Mathematics and Statistics, Newcastle University,
Newcastle upon Tyne, NE1 7RU, UK}
\author{Kandaswamy Subramanian}
\affiliation{Inter-University Centre for Astronomy and Astrophysics, Pune 411 007, India}

\begin{abstract}
We develop a new model of the fluctuation dynamo
in which the magnetic field is confined to thin flux ropes advected by
a multi-scale model of turbulence. Magnetic dissipation occurs only via
reconnection of the flux ropes.
This model can be viewed as an implementation of the asymptotic limit
$\Rm\to\infty$ for a continuous magnetic field, where magnetic dissipation
is strongly localized to small regions of strong field gradients.
We investigate the kinetic energy release into heat, mediated by the dynamo
action, both in our model and by solving the induction equation with the same
flow. We find that a flux rope dynamo is an order of magnitude more efficient
at converting mechanical energy into heat. The probability density of the
magnetic energy release in reconnections has a power-law form with the slope
$-3$, consistent with the Solar corona heating by nanoflares.
\end{abstract}

%47.35.Tv Magnetohydrodynamic waves
%47.65.Md Plasma dynamos
%95.30.Qd Magnetohydrodynamics and plasmas in astrophysics
%52.35.Vd Magnetic reconnection
%95.30.Qd Magnetohydrodynamics and plasmas
%96.60.Iv Magnetic reconnection  (Solar physics)
%96.60.qe Flares
%\keywords{}

\maketitle

%\section{Introduction}
The dynamo action, i.e., the amplification of magnetic field by the motion of
an electrically conducting fluid (plasma), is the most likely explanation for
astrophysical magnetic fields. Evolution of magnetic field $\mathbf{B}$
embedded in a flow  at a velocity $\mathbf{u}$ is governed by
\begin{equation}\label{induction}
  \pderiv{\mathbf{B}}{t}=\nabla \times (\mathbf{u} \times \mathbf{B})
        +\widehat{\cal L}\mathbf{B},
\end{equation}
where $\widehat{\cal L}$ is an operator describing magnetic dissipation. In
rarefied plasmas, such as the Solar corona, hot gas in spiral and elliptical
galaxies, galactic and accretion disc halos, and laboratory plasmas, an
important (if not dominant) mechanism for the dissipation of magnetic field is
the reconnection of magnetic lines rather than magnetic diffusion
\cite{priest:2000}, the latter modeled with $\widehat{\cal L}=\eta\nabla^2$
(if $\eta=\mbox{const}$). Discussions of dynamos often refer to magnetic
reconnection, but attempts to include any features specific of magnetic
reconnection to dynamo models are very rare \cite{B96}. On the other hand,
theories of magnetic reconnection rarely, if ever, refer to the dynamo action
as a mechanism maintaining magnetic fields. This paper attempts to bridge the
gap between the two major areas of magnetohydrodynamics by developing a dynamo
model explicitly incorporating magnetic reconnections.

The nature of the dissipation mechanism is important for the dynamo action.
For example, dynamo action with hyperdiffusion, $\widehat{\cal
L}=-\eta_1\nabla^4$ (and with a helical $\mathbf{u}$) has larger growth rate
and stronger steady-state magnetic fields than a similar dynamo based on
normal diffusion \cite{Brandenburg:2002}. This is not surprising as the
hyperdiffusion operator, having the Fourier dependence of $k^4$, rather than
 $k^2$ of the normal diffusion, has weaker magnetic dissipation at larger
scales. The release of magnetic energy in smaller regions (and larger current
densities) in hyperdiffusive dynamos may also lead to a higher rate of
conversion of kinetic energy to heat via magnetic energy. Magnetic
hyperdiffusion also appears in the context of continuous models of
self-organized criticality in application to the heating of the Solar corona
\cite{SOC}. The aim of such models is to reproduce the observed frequency
distribution of various flare energy diagnostics.

Magnetic reconnections may have an even more extreme form of the dissipation
operator than the hyperdiffusion: here magnetic fields  dissipate only when in
close contact with each other, so that the Fourier transform of $\widehat{\cal
L}$ can be expected to be negligible at all scales exceeding a certain
reconnection length $d_0$. It is then natural to expect that dynamos based on
reconnections (as opposed to those involving magnetic diffusion) will exhibit
faster growth of magnetic field, more intermittent spatial distribution and
stronger plasma heating. In this paper we consider dynamo action based on
direct modeling of magnetic reconnection. For this purpose, we follow the
evolution of individual closed magnetic loops in a model of turbulent flow
(known to be a dynamo) and reconnect them directly whenever their segments
come into sufficiently close contact, with appropriate magnetic field
directions. As we show here, our model exhibits a power-law probability
distribution of the magnetic energy release similar to that observed in the
Solar corona.

Magnetic reconnection is usually modeled with the induction equation,
$\widehat{\cal L}=\eta\nabla^2$ (perhaps including the Hall current),
and magnetic dissipation is enhanced due to the development of small-scale
motions and magnetic fields. This approach may or may not apply to magnetic
fields concentrated into flux ropes, where magnetic energy losses are strongly
reduced at large scales and, hence, more energy can be deposited at the
smaller scale of order the tube radius, where reconnections occur. Our model
explores this possibility. Furthermore, our model can be viewed as a numerical
implementation of the limit $\Rm\to\infty$ for a \textit{continuous
magnetic field}, where magnetic dissipation is confined to strongly localized
regions with exceptionally high magnetic field gradients.

\begin{figure}
  \begin{center}
    \includegraphics[width=0.25\textwidth, height=0.17\textwidth]{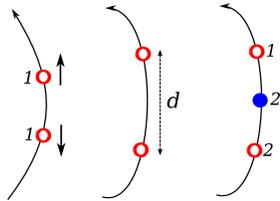}
    \caption{\label{fig1}
%(Color online)
The algorithm for inserting new trace particles in a stretched magnetic flux
tube. If the distance between any two particles (shown with
% red/
open circles) exceeds a length scale $d$, a new one is inserted between
them, shown with a
% blue/
filled circle. Labels represent magnetic field strength.
}
  \end{center}
\end{figure}

We model the evolution of thin flux tubes, frozen into a flow, each with
constant magnetic flux $\psi$. In this paper, we focus on the kinematic
behavior, where the velocity field is independent of magnetic field. To ensure
that $\nabla \cdot \mathbf{B}=0$, we require that our tubes always take the
form of closed loops. Numerically, we discretize the loops into fluid
particles and track their position and relative order (i.e., magnetic field
direction) by introducing a flag denoted $P$, with $P$ increasing along a
given magnetic flux tube. Initially the particles are set a small distance
apart, $0.75d$, where $d$ is an arbitrary (small) constant length scale. If,
during the evolution of the loops, the distance between neighboring fluid
particles on a loop becomes larger than $d$, we introduce a new particle
between them, as illustrated in Fig.~\ref{fig1}. We use linear interpolation
to place the new particle halfway between the old ones. The new separation
between the particles is thus greater than $0.5d$ -- this will be important
when we consider removing particles. Thus, the spatial resolution of our model
is $d$.

%% FIGURE 1--------------------------------------------------------------
%----------------------------------------------------------------------

Each particle is also assigned a flag $B$ (Fig.~\ref{fig1}) for the strength
of magnetic field at that point on the loop. Assuming magnetic flux
conservation and incompressability, magnetic field strength in the flux tube
is proportional to its length. Magnetic field is initially constant at all
particles, $B=1$. When a new particle is introduced, magnetic field is
doubled, as shown in Fig.~\ref{fig1}, at two out of three particles involved:
this prescription emerged from our experimentation with various schemes and
allows us to reproduce the evolution of magnetic field strength in a shear
flow. Conversely, when the flow reduces the separation of particles to less
than $0.5d$, we remove a particle. The value of the magnetic field strength
flag is also halved on the remaining particles in a manner consistent with the
above algorithm. We have verified that this prescription reproduces accurately
an exact solution of the induction equation for a simple shear flow.

%% FIGURE 4--------------------------------------------------------------
\begin{figure}
  \begin{center}
    \includegraphics[width=0.25\textwidth, height=0.17\textwidth]{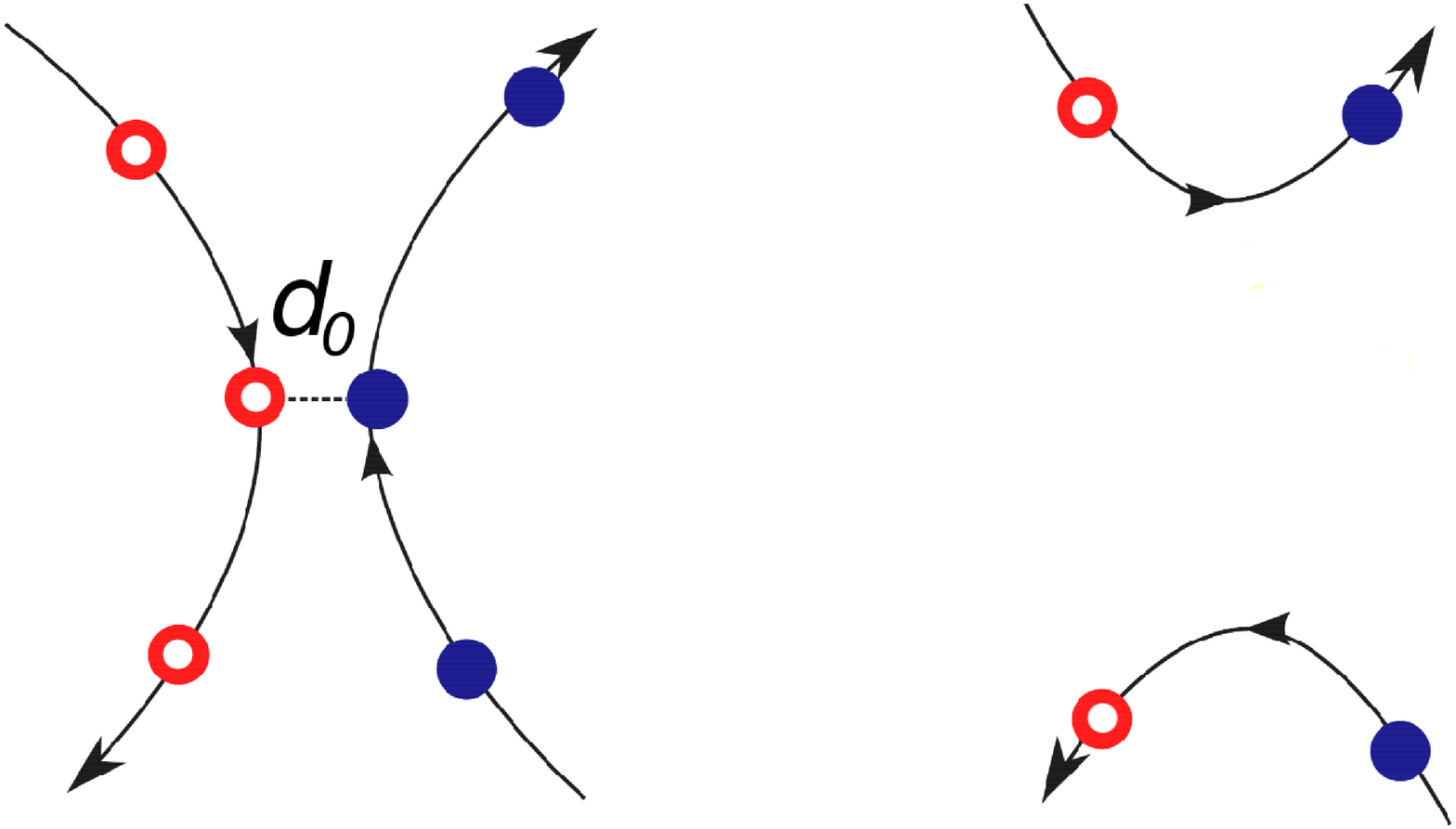}
    \caption{\label{fig4}
% (Color online)
Reconnection occurs when the distance between two trace particles reduces to
$d_0$ (left); the connection of the particles on a magnetic flux tube changes
after the reconnection, and the two closest particles are removed (right).}
\end{center}
\end{figure}
%----------------------------------------------------------------------

If the separation between two particles, which are not neighbours, becomes
less than a certain scale $d_0$, we reconnect their flux tubes by reassigning
the flags $P$ (Fig.~\ref{fig4}) which identify the particles ahead and those
behind of those involved in the reconnection. (To obtain meaningful numerical
results, $d_0$ has to be comparable to $d$, e.g., $d_0=1.5d$.) Two particles
are removed from the system after each reconnection event (and their magnetic
energy is lost, presumably to heat). We also monitor the cross product of
magnetic fields close to the reconnection point. By ensuring that its
magnitude is smaller than some tolerance $\epsilon\approx 10^{-2}$ and that
the magnetic fields in the reconnecting loops are (almost) oppositely
directed, we prevent parallel flux tubes with the same field direction from
reconnecting. We monitor the amount of magnetic energy released in each
reconnection event. To place the reconnection-based dynamo into a proper
perspective, we compare it with a dynamo obtained for the same velocity field,
but by solving the induction equation, i.e., Eq.~(\ref{induction}) with
$\widehat{\cal L}=\eta\nabla^2$. In particular, we compare the rates of
magnetic energy dissipation, which can be identified with the plasma heating
rate. We assume that the part of the magnetic energy which drives plasma
motion at a reconnection site (such as jets) is eventually dissipated into
heat as well, so that we consider that the whole magnetic energy released is
converted into heat. For the induction equation, the relevant quantity is
\begin{equation}
\gamma_\mathrm{i}=
{d\ln M}/{dt}
=\int_V \eta \mathbf{B} \cdot \nabla^2 \mathbf{B}\, dV\,\left[\int_V \mathbf{B}^2\, dV\right]^{-1}\;,
\end{equation}
where $M$ is the total magnetic energy. A similar quantity can be obtained for
the reconnection-based dynamo by adding the contributions of all reconnection
events to the magnetic energy release:
\begin{equation}
\gamma_\mathrm{r}=\frac{d\ln M}{dt}=
{\frac{1}{8\pi M\tau}\displaystyle \sum_{i=1}^{N_\tau}B_i^2S_iL_i}\;,
\end{equation}
where $\tau$ is a time interval during which $N_\tau$ reconnections occur (we
take $\tau$ to be equal to ten time steps; individual reconnection events
occur in a single time step), and $B_i$, $S_i$ and $L_i$ are the magnetic
field strength, the cross-sectional area and length of the reconnected (and
thus removed) flux tube segment associated with a particle number $i$. From
our assumption of frozen flux, $B_i S_i=\psi=\mbox{const}$, the total magnetic
energy $M$ is,
\begin{equation}
M=\displaystyle \sum_{i=1}^{N_\mathrm{tot}}\frac{B_i^2}{8\pi}S_iL_i
= \frac{\psi}{8\pi}\sum_{i=1}^{N_\mathrm{tot}}B_iL_i\;,
\end{equation}
where $N_\mathrm{tot}$ is the total number of particles, and
\begin{equation}
\gamma_\mathrm{r}={\tau}^{-1} {\sum_{i=1}^{N_\tau}B_iL_i}
        \left[{\sum_{i=1}^{N_\mathrm{tot}}B_iL_i}\right]^{-1}\;.
\end{equation}

Any comparison of the solutions of the induction equation with those from the
reconnection model is not straightforward because of the difference in the
control parameters of the two models: the magnetic Reynolds number
$R_\textrm{m}=u_0 l_0/\eta$ and the reconnection length $d_0$, respectively. A
proxy for the magnetic Reynolds number can be constructed from $d_0$ as
$\tilde{R}_\textrm{m}=u_0 l_0/(u_\textrm{r}d_0)$, where $u_\textrm{r}$ is the
characteristic reconnection speed. The reconnection-based dynamo is
significantly more efficient than the hydromagnetic dynamo, in the sense that
the growth rate of magnetic field in the former is significantly larger when
$R_\textrm{m}\approx\tilde{R}_\textrm{m}$. Therefore, in order to achieve
conservative conclusions, we compare dynamos with \textit{similar growth
rates\/} of magnetic field. Thus, $R_\textrm{m}>\tilde{R}_\textrm{m}$ in the
models compared. Magnetic field growth in a dynamo is obtained from the
difference between the magnetic stretching and dissipation rates. In the
reconnection-based dynamo, both are larger than those in a similar
diffusion-based dynamo, but their difference is kept the same in the models
which we compare below.

We consider dynamos driven by two types of flow. Firstly, this is the
Kinematic Simulation (KS) model of a turbulent flow \cite{Osborne:2006}, known
to be a dynamo \cite{Wilkin:2007}. Here velocity at a position $\mathbf{x}$
and time $t$ is
\begin{equation}\label{uF}
{\bf u}({\bf x},t)= \sum_{n=1}^{N}\left(\mathbf{A}_n \times {\bf k}_n
\cos\phi_n + {\bf B}_n \times {\bf k}_n \sin\phi_n \right),
\end{equation}
where $\phi_n=\mathbf{k}_n \cdot {\bf{x}} + \omega_n t$, $N$ is the number of
modes, $\mathbf{k}_n$ and $\omega_n=k_n u_n$ are their wave vectors and
frequencies. An advantage of using this flow is that the energy spectrum,
$E(k_n)$ is controllable via appropriate choice of $\mathbf{A}_n$ and
$\mathbf{B}_n$. We also note that $\nabla\cdot\mathbf{u}=0$. We adopt an
energy spectrum which reduces to $E(k)\propto k^{-p}$ for $1\ll k\ll k_N$,
with $k=1$ at the integral scale; $p=5/3$ produces the Kolmogorov spectrum,
and $k_N$ is the cut-off scale. We have adapted (\ref{uF}) to periodic
boundary conditions.

We also used the ABC flow of the form \cite{Childress:1995}
\begin{equation}
\mathbf{u}=(\cos y+\sin z,\sin x+\cos z, \cos x+\sin y)\;,
\end{equation}
also known to support dynamo action, to demonstrate that our results are not
sensitive to the form of the flow.

%----------------------------------------------------------------------------
\begin{figure}
  \begin{center}
    \psfrag{e}[c][position=1mm]{\vspace{1mm} $\gamma_\mathrm{i} l_0/u_0,\ \ \gamma_\mathrm{r} l_0/u_0$}
    \psfrag{t}[c][position=1mm]{\vspace{1mm} $t u_0/l_0$}
    \includegraphics[width=0.35\textwidth, height=0.25\textwidth]{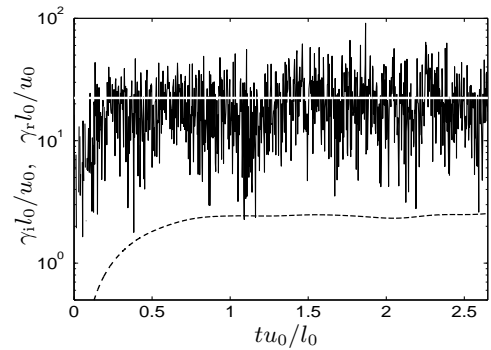}
    \caption{\label{fig5} Magnetic energy release rates from
the induction equation (dashed) and the flux rope model (solid). The former
has a mean value of 2.4 (here $\Rm=1200$) once the eigensolution has
developed. The latter (with $\tilde{R}_\textrm{m}=174$) has a mean value of 23
(thick horizontal line).}
\end{center}
\end{figure}
%--------------------------------------------------------------------

The initial condition is a random ensemble of closed magnetic loops, and both
the induction equation and the flux rope model are evolved with the same
velocity field (apart from the overall normalization to provide comparable
growth rates of magnetic field). The initial condition for the induction
equation is obtained by Gaussian smoothing of the magnetic field in the ropes
(this procedure preserves $\nabla\cdot\mathbf{B}=0$). To evolve the induction
equation, we use the Pencil Code \cite{PC:2002} on a $256^3$ mesh with
$1000<\Rm<1500$ in a periodic box. The test particles in the flux ropes are
evolved using a $4^{\textrm{th}}$ order Runge--Kutta scheme, with a time step
of $l_N/(20u_N)$. The algorithm for inserting and removing points is applied
every time step, and the reconnection algorithm, every ten time steps. We
choose $d$ to be 1/4 of the smallest length scale in the flow and set
$d_0/d=1.5$.

Figure~\ref{fig5}  shows the energy release rates in simulations where the
growth rate of the magnetic field is $\sigma=0.16$ in both simulations (with
the unit time $l_0/u_0$). The dashed line shows the energy release rate from a
simulation of induction equation with $\Rm=1200$, which has the mean energy
release rate $\gamma_\mathrm{i}\approx2.4$. The solid line shows the
corresponding results from the flux rope dynamo, with the mean value plotted
as a dashed horizontal line. The mean value of the energy release rate from
the reconnecting flux rope dynamo is $\gamma_\mathrm{r}\approx23$, an order of
magnitude larger. Also note strong fluctuations in the energy release rate
from the reconnection model, which are absent in the solutions of the
induction equation.

Dynamos with the ABC flow behave similarly. With $\Rm=55$, the induction
equation gives an energy release rate of about $\gamma_\mathrm{i}=0.6$. The
corresponding flux rope dynamo with the same growth rate ($0.02$) has the
energy release rate of $\gamma_\mathrm{i}\approx6.7$, again ten times larger.

Our approach is deliberately oversimplified with respect to the (incompletely
understood) physics of magnetic reconnection. Nevertheless, we can argue that
our model is conservative with respect to the reconnection efficiency. The
reconnecting segments of magnetic lines in our model approach each other at a
speed $u_\mathrm{r}\simeq u_0\Rey^{-1/4}$ for the Kolmogorov spectrum, equal
to velocity at the \textit{small\/} scale $d_0\ll l_0$ with $l_0$ the
energy-range scale of the flow and $d_0$ assumed to be close to the turbulent
cut-off scale. If magnetic field is strong enough, the Alfv\'en speed
$V_\mathrm{A}$, which controls magnetic reconnection in more realistic models,
is of order $u(l_0)$. Then $u_\mathrm{r}\ll V_\mathrm{A}$ and our model is
likely to underestimate the efficiency of reconnections. The Sweet--Parker
reconnection proceeds at a speed of order $V_\mathrm{A}\Rm^{-1/2}$, whereas
the Petschek reconnection speed is comparable to $V_\mathrm{A}/\ln\Rm$
\cite{priest:2000}. For $u_0\simeq V_\mathrm{A}$ and $\Rm\simeq\Rey\gg1$, the
reconnection rate in our model is larger than the former but much smaller than
the latter.

%%--------------------------------------------------------------
\begin{figure}
  \begin{center}
    \psfrag{z}[c]{\mbox{}$\log \Delta M/B_\mathrm{rms}^{2^{{\ }^{{\ }^{\ }}}}$}
    \psfrag{p}[c]{\mbox{}\vspace*{3em}Probability density}
    \includegraphics[height=0.2\textwidth]{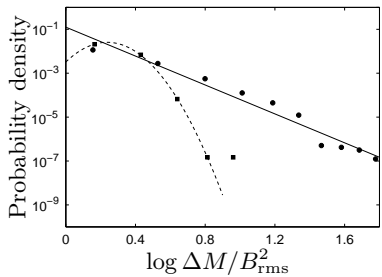}
    \caption{\label{pdf} Probability density for the normalized magnetic
energy release in individual reconnection events, $\Delta M/B_\mathrm{rms}^2$,
from the time series of Fig.~\protect\ref{fig5}, for the flux rope dynamo
(circles) and the diffusive dynamo with the same magnetic field growth rate
and velocity field form (squares). A power-law fit to the former and a
Gaussian fit to the latter are shown solid and dashed, respectively.}
\end{center}
\end{figure}
%----------------------------------------------------------------------

A remarkable feature of the energy release in the rope dynamo is that its
probability distribution has a power law as shown in Fig.~\ref{pdf},
$f(x)\propto x^{-s}$, where $x=\Delta M/B_\mathrm{rms}^2$ is the magnetic
energy released in a reconnection event normalized to the mean magnetic
energy, with the slope $s\approx3.3$. Importantly, the same scaling,
$s\approx3.0$, emerges when we use the ABC flow instead of KS. A similar
exponent arises in a reconnection model for the corona \cite{Hughes:2003}
where, however, dynamo action is not included. Thus, weak `flares' dominate
the energy release in our reconnection-based system, as in the nanoflare model
of coronal heating \cite{P83}. 
Interestingly more recent results with a nonlinear adaptation of the model \cite{Baggaley:2009} retains this feature with $s\approx -3.1$ for the KS flow in
the statistically steady state. 
We stress that the power-law behavior is not
related to the self-similar nature of the velocity field: solution of the
induction equation with the same velocity field, also shown in Fig.~\ref{pdf},
has an approximately Gaussian probability distribution. It is not as yet clear
if the flux rope dynamo represents a physical example of self-organized
criticality, but the system does possess some of the required properties. In
particular, our reconnection model has a natural threshold in terms of the
current density $J>B_\mathrm{min}/d_0$, where $B_\mathrm{min}=1$ is the
minimum magnetic field, and, as we argue above, our model can be viewed as an
extreme case of magnetic hyperdiffusivity. Furthermore, our simulations are
kinematic (so, magnetic energy density is assumed to be small), whereas the
Solar corona is magnetically dominated. The importance of this distinction
needs to be carefully investigated.

To summarize, we have confirmed that the dynamo action is sensitive to the
nature of magnetic dissipation and demonstrated that magnetic reconnections
(as opposed to magnetic diffusion) can significantly enhance the dynamo
action. We have explored the kinematic stage of the fluctuation dynamo in a
chaotic flow that models hydrodynamic turbulence and in the ABC flow, with the
only magnetic dissipation mechanism being the reconnection of magnetic lines
implemented in a direct manner. In our model, where magnetic dissipation is
suppressed at all scales exceeding a certain scale $d_0$, the growth rate of
magnetic field exceeds that of the fluctuation dynamo, based on magnetic
diffusion, with the same velocity field. Even when the velocity field of the
reconnection-based dynamo is reduced in magnitude as to achieve similar growth
rates of magnetic energy density, the rate of conversion of magnetic energy
into heat in the reconnection dynamo is a order of magnitude larger than in
the corresponding diffusion-based dynamo. Thus, reconnections more efficiently
convert the kinetic energy of the plasma flow into heat, in our case with the
mediation of the dynamo action. This result, here obtained for a kinematic
dynamo, can have serious implications for the heating of rarefied, hot plasmas
where magnetic reconnections dominate over magnetic diffusion (such as the
corona of the Sun and star, galaxies and accretion discs). In contrast to the
fluctuation dynamo based on magnetic diffusion, the probability distribution
function of the energy released in the flux rope dynamo has a power law form
not dissimilar to that observed for the Solar flares.

We are grateful to P.~H.~Diamond, R.~M.~Kulsrud, A.~Schekochihin and
A.~M.~Soward for useful discussions and suggestions. AS is grateful to IUCAA
for financial support and hospitality.

\bibliographystyle{pf}
\bibliography{my}

\begin{thebibliography}{10}

\bibitem{priest:2000}
E.~{Priest} and T.~{Forbes},
\newblock {\em {Magnetic Reconnection}},
\newblock Cambridge University Press, 2000.

\bibitem{B96}
E.~G. {Blackman},
\newblock Phys.\ Rev.\ Lett. {\bf 77}, 2694 (1996).

\bibitem{Brandenburg:2002}
A.~Brandenburg and G.~R. Sarson,
\newblock Phys.\ Rev.\ Lett. {\bf 88}, 055003 (2002).

\bibitem{SOC}
P.~{Charbonneau}, S.~W. {et al.}~%and {McIntosh}, H.-L. {Liu}, and T.~J.
  %{Bogdan},
\newblock Solar Phys. {\bf 203}, 321 (2001).

\bibitem{Osborne:2006}
D.~Osborne et~al.,
\newblock Phys.~Rev.~E {\bf 74}, 036309 (2006).

\bibitem{Wilkin:2007}
S.~L. Wilkin et~al.,
\newblock Phys.~Rev.~Lett. {\bf 99}, 134501 (2007).

\bibitem{Childress:1995}
S.~Childress and A.~Gilbert,
\newblock {\em Stretch, Twist, Fold: The Fast Dynamo},
\newblock Springer, Berlin, 1995.

\bibitem{PC:2002}
A.~{Brandenburg},
\newblock Comp.\ Phys.\ Comm. {\bf 147}, 471 (2002).

\bibitem{Hughes:2003}
D.~Hughes et~al.,
\newblock Phys.~Rev.~Lett. {\bf 90}, 131101 (2003).

\bibitem{P83}
E.~N. {Parker},
\newblock \apj {\bf 264}, 642 (1983).

\bibitem{Baggaley:2009}
A.~W. {Baggaley}, C.~F. {Barenghi}, A.~{Shukurov}, and K.~{Subramanian},
\newblock Astron. Nachr {\bf 331} (2010).

\end{thebibliography}

\end{document}